\documentclass[10pt,twocolumn,letterpaper]{article}

\usepackage{wacv}              % To produce the CAMERA-READY version
\usepackage{comment}
\usepackage{caption}
\usepackage{subcaption}
\usepackage{pbalance}
\definecolor{wacvblue}{rgb}{0.21,0.49,0.74}
\usepackage[pagebackref,breaklinks,colorlinks,allcolors=wacvblue]{hyperref}
\usepackage[accsupp]{axessibility}  % Improves PDF readability for those with disabilities.
\usepackage{xcolor} 
\usepackage{balance}

\def\wacvPaperID{1654} % *** Enter the WACV Paper ID here
\def\confName{WACV}
\def\confYear{2026}

\newcommand{\sys}{SpeechQualityLLM\xspace}
\title{\sys: LLM-Based Multimodal Assessment of Speech Quality}

\author{Mahathir Monjur\\
UNC Chapel Hill\\
{\tt\small mahathir@cs.unc.edu} 
\and
Shahriar Nirjon\\
UNC Chapel Hill\\
{\tt\small nirjon@cs.unc.edu}
}

\begin{document}
\maketitle

%% Abstract %%
\begin{abstract}
Objective speech quality assessment is central to telephony, VoIP, and streaming systems, where large volumes of degraded audio must be monitored and optimized at scale. Classical metrics such as PESQ and POLQA approximate human mean opinion scores (MOS) but require carefully controlled conditions and expensive listening tests, while learning-based models such as NISQA regress MOS and multiple perceptual dimensions from waveforms or spectrograms, achieving high correlation with subjective ratings yet remaining rigid: they yield fixed scalar scores, do not support interactive, natural-language queries, and do not natively provide textual rationales. In this work, we introduce \emph{\sys}, a multimodal speech quality question--answering (QA) system that couples an audio encoder with a language model and is trained on the NISQA corpus using template-based question--answer pairs covering overall MOS and four perceptual dimensions (noisiness, coloration, discontinuity, and loudness) in both single-ended (degraded only) and double-ended (degraded plus clean reference) setups. Instead of directly regressing scores, \sys is supervised to generate textual answers from which numeric predictions are parsed and evaluated with standard regression and ranking metrics; on held-out NISQA clips, the double-ended model attains a MOS mean absolute error (MAE) of approximately 0.41 with Pearson correlation of 0.86, with competitive performance on dimension-wise tasks. Beyond these quantitative gains, \sys offers a flexible natural-language interface in which the language model acts as an audio quality expert: practitioners can query arbitrary aspects of degradations, prompt the model to emulate different listener profiles to capture human variability and produce diverse but plausible judgments rather than a single deterministic score, and thereby reduce reliance on large-scale crowdsourced tests and their monetary cost. We provide a general pipeline for adapting large language models to specialized audio quality assessment tasks via lightweight multimodal alignment. Code, model weights, and experimental results are available at \href{https://github.com/Monjur-Mahathir/Speech-Quality-LLM}{GitHub}.
\end{abstract}   

%% Abstract %%

\section{Introduction}

Speech quality assessment is a core component of modern audio communication systems, including telephony, VoIP, video conferencing, and media streaming~\cite{sq1, sq2, sq3}. Service providers must continuously monitor and optimize the perceived quality of speech signals across highly variable networks, devices, and acoustic environments. When quality degrades---due to noise, compression artifacts, packet loss, reverberation, or other impairments---user experience suffers, leading to dropped calls, reduced engagement, and churn. As a result, scalable, reliable, and interpretable speech quality assessment methods are essential not only for offline codec and algorithm design, but also for real-time monitoring, troubleshooting, and closed-loop optimization of large-scale communication platforms.

\begin{figure}[t]
    \centering
    \includegraphics[width=0.65\linewidth]{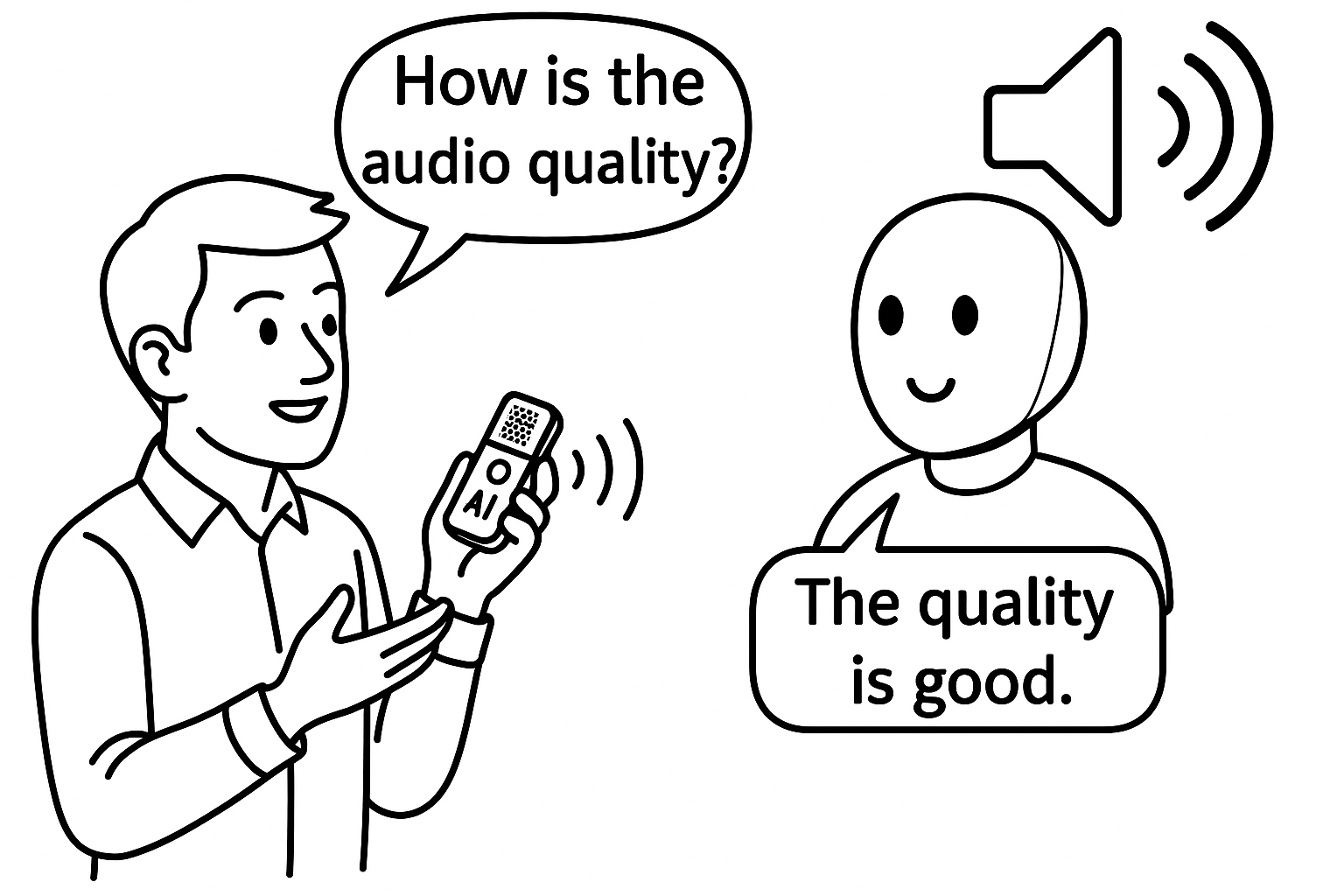}
    \caption{Illustration of our use-case: a user records speech on a device, then queries the AI “quality expert’’ that listens to the audio and provides an interpretable assessment of overall quality and specific artifacts.}
    \label{fig:intro}
    \vspace{-10pt}
\end{figure}

In practice, multiple metrics are used to quantify quality along different axes. Objective measures such as PESQ~\cite{pesq}, POLQA~\cite{polqa}, STOI~\cite{stoi}, SI-SDR~\cite{sisdr}, or segmental SNR estimate aspects of fidelity, intelligibility, and distortion by comparing a degraded signal to a clean reference or by analyzing the degraded signal alone. At the same time, practitioners are often interested in decomposed dimensions such as noisiness, coloration, discontinuity, and loudness. Despite this variety, the dominant target for system optimization remains the mean opinion score (MOS), where human listeners rate the overall quality of a clip on a discrete scale (typically 1--5), and scores are averaged across listeners. This has become the de facto ground truth for codec tuning, benchmarking, and service-level agreements, because it reflects holistic user satisfaction rather than any single low-level artifact.

However, obtaining MOS at scale is costly and operationally complex. Conventional subjective tests, such as those recommended by ITU-T~\cite{itu1,itu2,itu3}, require carefully controlled listening conditions, calibrated playback equipment, and trained subjects, making them expensive and slow to run. Crowdsourcing alleviates some of this burden but introduces its own challenges: rater reliability must be monitored, screening procedures and gold questions are needed, and platform effects (e.g., headphones vs.\ laptop speakers, noisy environments) introduce additional variance. For large datasets, the cost of collecting high-quality MOS labels grows linearly with the number of clips and the number of raters per clip, which is prohibitive when providers need to evaluate millions of utterances or continuously monitor live traffic. As a result, subjective tests are typically reserved for periodic benchmarking campaigns rather than integrated into routine monitoring loops.

To address scalability, the community has developed learning-based MOS predictors that approximate human judgments directly from audio signals. Given large datasets of degraded waveforms or spectrograms paired with MOS (or dimension-wise ratings), neural networks are trained with regression and ranking objectives to match these human scores. Systems like NISQA~\cite{nisqa} and DNSMOS~\cite{dnsmos} achieve high correlation with subjective ratings on in-domain test sets and can be deployed as fast, fully automatic predictors. Yet these models remain fundamentally rigid: they map each input to a single deterministic scalar (or a small set of scalars), without any notion of uncertainty, listener variability, or user intent, and they provide no textual explanations of their decisions. Moreover, because MOS itself is a noisy approximation of human opinion—shaped by listener profile, context, and task—models trained on one dataset often generalize poorly to new domains, languages, or device conditions where the rater population and usage scenarios differ.

In this work, we explore an alternative perspective: treating speech quality assessment as an interactive question--answering task and using a large language model (LLM) as the \emph{expert listener}. Our system, \emph{\sys}, couples an audio encoder~(either Audio Spectrogram Transformer AST~\cite{ast} or Whisper~\cite{whisper}) with a text-based LLM (Llama 3.1-8B~\cite{lama31}) model and is trained to answer natural-language questions about overall MOS and specific perceptual dimensions. Because LLMs are inherently conditioned on prompts, we can explicitly specify different listener profiles or preferences (e.g., a user highly sensitive to background noise, or a professional who prioritizes speech clarity over loudness), and the model can adapt its qualitative and quantitative judgments accordingly. This makes it possible to simulate a diverse panel of virtual listeners and capture the inherent randomness and subjectivity in human perception, instead of collapsing everything into a single global mapping. At the same time, the system supports rich, text-based interactions: practitioners can request justifications, or query particular aspects of degradations, obtaining interpretable explanations rather than opaque scalar outputs.

Building on this formulation, our contributions are as follows:
\begin{itemize}

\item We propose a general pipeline for adapting LLMs to speech quality assessment: we construct methodologically controlled question--answer pairs from an existing MOS dataset (NISQA~\cite{nisqa}), covering both overall and dimension-wise ratings, and use them to align an audio encoder with a pretrained LLM via lightweight multimodal adapters. 

\item We instantiate this pipeline in \emph{\sys} and show that it achieves competitive predictive performance on held-out NISQA clips, with double-ended models reaching a MOS mean absolute error of approximately 0.41 and Pearson correlation of 0.86, while also supporting both single-ended and double-ended configurations. 

\item We demonstrate qualitative advantages that are difficult to obtain with conventional models: \sys can emulate different listener profiles, produce diverse yet plausible judgments that reflect human variability, and provide explanation-rich responses to arbitrary queries about degradations, thereby reducing reliance on large-scale crowdsourced listening tests and their associated cost. 
\end{itemize}

Together, these results suggest that LLM-based expert listeners offer a promising path toward scalable, interactive, and human-centered speech quality assessment.
\section{Background}

Objective speech quality assessment has long relied on a combination of subjective listening tests and hand-crafted objective metrics. In subjective tests, human listeners assign mean opinion scores (MOS) to speech clips on an ordinal scale. These scores are treated as the gold standard for evaluating codecs, noise suppressors, and end-to-end communication pipelines, and form the basis of recommendations such as ITU-T P.800~\cite{itu1}, P.808~\cite{itu2}, and related protocols~\cite{itu3} for controlled laboratory and crowdsourced studies. On the objective side, metrics such as PESQ~\cite{pesq}, POLQA~\cite{polqa}, STOI~\cite{stoi}, and variants thereof attempt to predict these MOS ratings from signal comparisons or intrinsic features, and are widely adopted in telephony, VoIP, and conferencing systems as proxies for real user perception. Despite their success, these tools are usually designed for specific bandwidths, languages, or device conditions, and require careful calibration when moved to new domains. However, both subjective MOS tests and traditional objective metrics are costly, inflexible, and offer little interpretability, whereas \sys targets scalable, interactive, and explanation-rich judgments driven by a language model acting as an expert listener.

To reduce the dependence on repeated listening tests, recent work has turned to deep neural networks that approximate MOS and related perceptual dimensions directly from audio. Non-intrusive and intrusive models are trained on large labeled corpora, mapping waveforms or spectrograms to overall MOS. Systems like NISQA~\cite{nisqa} and DNSMOS~\cite{dnsmos} have shown that convolutional and attention-based architectures can achieve high correlation with human scores on in-domain test sets, and have become attractive for real-time monitoring in cloud and edge deployments. In practice, they have become de facto backbones for benchmarking enhancement algorithms and tuning production pipelines. However, these deep models still reduce each input to a small set of fixed scores with no notion of listener profile, variability, or intent, while our system uses a language model interface to expose nuanced, profile-aware, and conversational assessments of quality.

In parallel, the emergence of large language models tailored to speech has opened new possibilities for multimodal understanding. Speech-centric LLMs and audio-augmented conversational models can transcribe, translate, and summarize speech, answer content questions about audio clips, and integrate prosodic cues into downstream reasoning~\cite{audiollm1, audiollm2}. By combining powerful sequence modeling with instruction-tuned dialogue capabilities, these models are increasingly used as general-purpose audio assistants that can interpret complex prompts and generate rich natural language responses grounded in acoustic evidence. Recent multimodal extensions further allow these systems to attend jointly to text, audio, and sometimes visual signals, bringing them closer to human-like conversational agents for spoken content. This trend suggests that language models can serve not only as transcription engines but also as high-level perceptual and analytical tools for spoken content. However, existing speech LLMs rarely treat quality assessment as a first-class task, whereas our system explicitly aligns an audio encoder with an LLM to act as an interactive quality expert that can be queried about MOS, dimension-wise ratings, and listener-dependent preferences.

\begin{figure*}[t]
    \centering
    \includegraphics[width=0.8\linewidth]{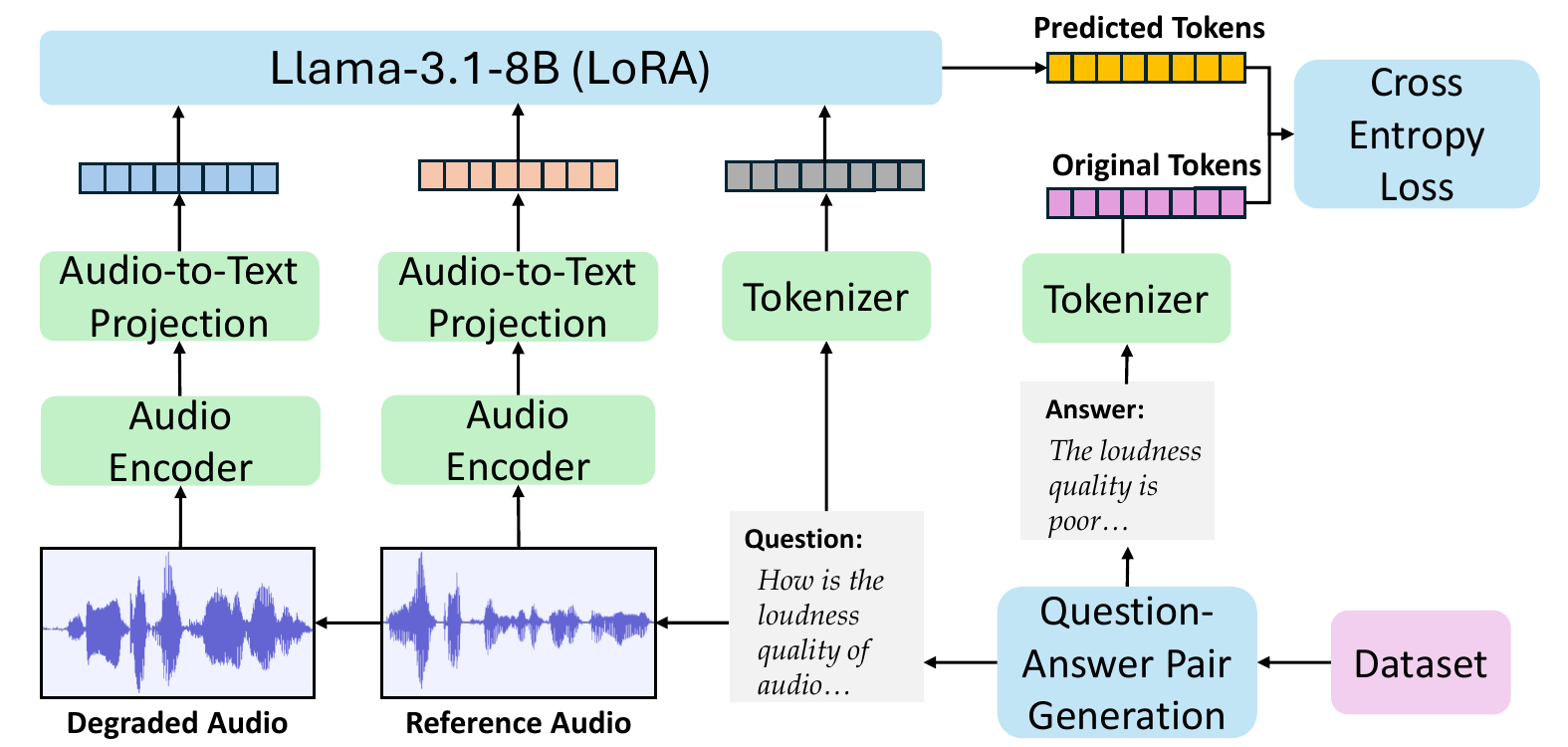}
    \caption{\textbf{\sys architecture.} Degraded and (optionally) reference speech are encoded by an AST or Whisper audio encoder, pooled and projected into a fixed number of audio tokens, and concatenated with question tokens from the QA template. The resulting multimodal sequence is fed to a LoRA-tuned LlaMA decoder, which autoregressively generates an answer. During training, predicted answer tokens are compared against the ground-truth answer tokens with a cross-entropy loss. In the single-ended setting, the reference branch is omitted.}
    \label{fig:system_diagram}
\end{figure*}

\section{Method}

\subsection{Question--Answer Dataset Generation}

We construct our training data from the NISQA corpus, which provides degraded speech clips with metadata including database split, corresponding reference signals, and listener scores for overall MOS and four perceptual dimensions (noisiness, coloration, discontinuity, loudness) along with per-dimension standard deviations. For each sample, we generate a short textual question and answer on the fly rather than relying on a fixed, precomputed prompt set. We maintain a template bank with multiple natural-language question–answer templates per task family; at each dataset index, we (i) sample a task type uniformly from the allowed families, (ii) select a template pair for that task, and (iii) fill in its placeholders using the ground-truth MOS and dimension-wise scores. This yields diverse yet label-consistent QA pairs for the same underlying audio and works in both full-reference and no-reference settings. Table~\ref{tab:qa-tasks} summarizes the task families, and Figure~\ref{fig:system_diagram} illustrates the overall system architecture.

For single-ended experiments, we operate solely on the degraded waveform. Each file is loaded, converted to mono if necessary, resampled to $16$\,kHz, and normalized in duration to a fixed $10$\,s window by random cropping for longer utterances and zero-padding for shorter ones. The resulting segment is transformed into log-mel time--frequency features using a pretrained audio front-end (AST or Whisper), which then serves as the input representation to the audio encoder. In the double-ended setting, we further require the corresponding clean reference signal. Both degraded and reference waveforms are loaded, resampled to a common sampling rate, and time-aligned by estimating their relative delay via cross-correlation and shifting accordingly. After alignment, we crop both signals to a shared segment of at most $10$\,s and pad as needed, before extracting features with the same AST or Whisper front-end, depending on the backbone used in the particular configuration. In all cases, each dataset entry consists of the processed audio features, a natural-language question--answer pair generated from the NISQA labels, and a structured set of ground-truth scores (overall MOS, four perceptual dimensions, and their standard deviations) that we later use for quantitative evaluation.

\begin{table*}[t]
  \centering
  \small
  \vspace{5pt}
  \caption{Families of question--answer templates used to train the model.}
  \begin{tabular}{|p{1.7cm}|p{4.0cm}|p{4.5cm}|p{5.8cm}|}
    \toprule
    Task family & Target labels & Question type & Answer format \\
    \midrule
    MOS-numeric & Overall MOS & Ask for overall quality on 1--5 scale & Single numeric MOS (possibly embedded in a short sentence) \\
    \hline
    Dim-numeric & One of \{noise, coloration, discontinuity, loudness\} & Ask for quality of a specific dimension on 1--5 scale & Single numeric score for that dimension \\
    \hline
    Dim-categorical & Same four dimensions (binned) & Ask for a descriptive rating of a dimension & Verbal category (\emph{very bad}--\emph{excellent}) mapped from the numeric score \\
    \hline
    Multi-dim & MOS + all four dimensions & Ask for a joint assessment across dimensions & Short summary sentence that mentions MOS and all four dimension scores \\
    \hline
    Explanatory & Overall MOS (and dimensions) & Ask for an overall quality judgment and explanation & Short natural-language justification that includes an MOS and mentions salient artifacts or dimensions \\
    \bottomrule
  \end{tabular}
  \label{tab:qa-tasks}
\end{table*}

\subsection{Model Architecture}

\sys couples an audio encoder with a large language model (LLM) via a lightweight projection layer that converts audio features into a sequence of pseudo-token embeddings. Let $x_{\mathrm{deg}}$ and $x_{\mathrm{ref}}$ denote the degraded and (optional) reference waveforms after resampling and cropping. For the double-ended setting, we first compute frame-level audio representations:
\[
  h_{\mathrm{deg}} = g(x_{\mathrm{deg}}) \in \mathbb{R}^{T_{\mathrm{deg}} \times d_a},
  \quad
  h_{\mathrm{ref}} = g(x_{\mathrm{ref}}) \in \mathbb{R}^{T_{\mathrm{ref}} \times d_a},
\]
where $g(\cdot)$ is either an AST backbone (with $d_a = 768$) or a Whisper encoder (with $d_a = 1280$). We then apply an \emph{audio projection layer} that pools these variable-length sequences to a fixed token budget and maps them to the LLM embedding space. The projection layer consists of temporal adaptive average pooling to a fixed length $L_a$ (typically $L_a = 128$), followed by layer normalization and a linear projection to dimension $d_t$, the hidden size of the LLM (for the Llama 3.1 8B model, $d_t = 4096$). This yields:
\[
  z_{\mathrm{deg}} = \phi(h_{\mathrm{deg}}) \in \mathbb{R}^{L_a \times d_t}, \qquad
  z_{\mathrm{ref}} = \phi(h_{\mathrm{ref}}) \in \mathbb{R}^{L_a \times d_t},
\]
which can be treated as contiguous blocks of ``audio tokens'' in the LLM input. In the single-ended variant, we only instantiate $z_{\mathrm{deg}}$ and omit $z_{\mathrm{ref}}$.

Textual components are handled by a LlaMA-family causal LLM. We tokenize three segments: a system and user prompt $P$ that includes the question, a short delimiter segment $E$ that marks the start of the assistant response, and the answer sequence $Y$. Let $e(\cdot)$ be the LLM token embedding function. The final input sequence of embeddings for a double-ended example is
\[
  \underbrace{e(P)}_{\text{prompt}}
  \;\Vert\;
  \underbrace{z_{\mathrm{deg}}}_{\text{degraded audio}}
  \;\Vert\;
  \underbrace{z_{\mathrm{ref}}}_{\text{reference audio}}
  \;\Vert\;
  \underbrace{e(E)}_{\text{delimiter}}
  \;\Vert\;
  \underbrace{e(Y)}_{\text{answer}},
\]
where $\Vert$ denotes concatenation along the sequence dimension. The attention mask is constructed analogously by concatenating the prompt mask, unit masks for audio tokens, the delimiter mask, and the answer mask; for single-ended inputs, we drop $z_{\mathrm{ref}}$ and its mask. At training time, the full sequence is passed through the LLM; at inference time, we only feed the prefix $[e(P) \Vert z_{\mathrm{deg}} \Vert z_{\mathrm{ref}} \Vert e(E)]$ and let the LLM autoregressively generate the answer tokens.

To adapt the LLM efficiently, we use low-rank adaptation (LoRA) on the query and key projections of all attention layers while keeping the base LlaMA weights in 4-bit quantized form. The audio encoders are kept frozen or lightly fine-tuned (e.g., by unfreezing a subset of attention query parameters in the AST backbone), and the only fully trainable components on the audio side are the projection layers that map encoder features into the LLM embedding space. This design keeps the number of trainable parameters modest while allowing the model to learn a shared multimodal representation that aligns audio quality cues with natural-language reasoning.

\subsection{Training and Output Generation}

We train separate models for single-ended and double-ended configurations, and for each setting we instantiate variants with either an AST-based or a Whisper-based audio encoder. For every configuration, we follow the official NISQA partitioning to derive training, validation, and test splits, and construct the corresponding QA dataset with a fixed random seed and the five task families in Table~\ref{tab:qa-tasks}. Mini-batches are assembled by jointly padding audio features and text sequences: audio representations are zero-padded along the temporal dimension, while the tokenized prompt, delimiter, and answer segments are left-padded so that end-of-sequence positions are aligned across examples. Models are optimized with AdamW using a batch size of $4$, a learning rate of $4\times 10^{-5}$, weight decay of $0.01$, and a linearly warmed-up learning-rate schedule. Training proceeds for $10{,}000$ optimization steps over the training data, and we monitor validation loss throughout to select the final checkpoint for each configuration.

The training objective is a standard next-token cross-entropy loss applied only to the answer segment. Let $Y=(y_1,\dots,y_T)$ be the tokenized answer, and let $\mathrm{prefix}$ denote the concatenation of prompt, audio tokens, and delimiter. Given model parameters $\theta$, the LLM defines a conditional distribution $p_\theta(y_t \mid \mathrm{prefix}, y_{<t})$ for each answer position. We compute the loss
\[
  \mathcal{L}(\theta)
  = - \frac{1}{T} \sum_{t=1}^{T}
    \log p_\theta\bigl(y_t \mid \mathrm{prefix}, y_{<t}\bigr),
\]
by slicing the LLM logits to exclude positions corresponding to the prompt, audio tokens, and delimiter, and aligning them with the shifted answer tokens. Gradients are propagated through the LoRA adapters, the audio projection layers, and any unfrozen audio encoder parameters. 

At inference time, we follow the same preprocessing pipeline to obtain audio features and construct the prompt and delimiter segments. We then compute the multimodal prefix embeddings, feed them into the LLM's generation API, and decode the resulting tokens into natural language. Numeric MOS and dimension-wise predictions are extracted from the generated text using simple regular expressions and rule-based mappings from adjectives to score bins, enabling direct comparison to ground-truth MOS and dimension ratings using standard regression and ranking metrics. Crucially, because the core output is free-form text, users can also modify the prompt to instruct the model to emulate different listener profiles (e.g., a user who is highly sensitive to background noise versus one who primarily cares about loudness), thereby introducing controlled variability in the judgments and reducing the need for repeated large-scale crowdsourced listening tests.

\begin{figure*}[t]
  \centering
  % First image
  \begin{subfigure}[b]{0.2\textwidth}
    \centering
    \includegraphics[width=\linewidth]{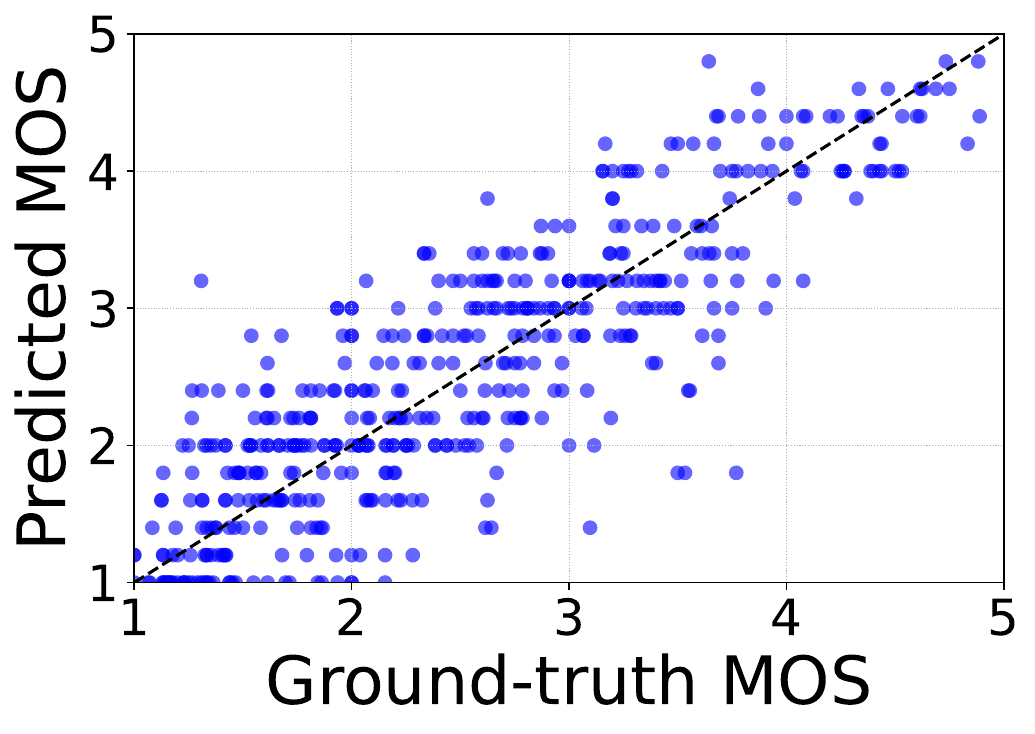}
    \caption{Full-ref AST (finetuned)}
    \label{fig:mos-scatter-fullref-ast-ft}
  \end{subfigure}
  % Second image
  \begin{subfigure}[b]{0.19\textwidth}
    \centering
    \includegraphics[width=\linewidth]{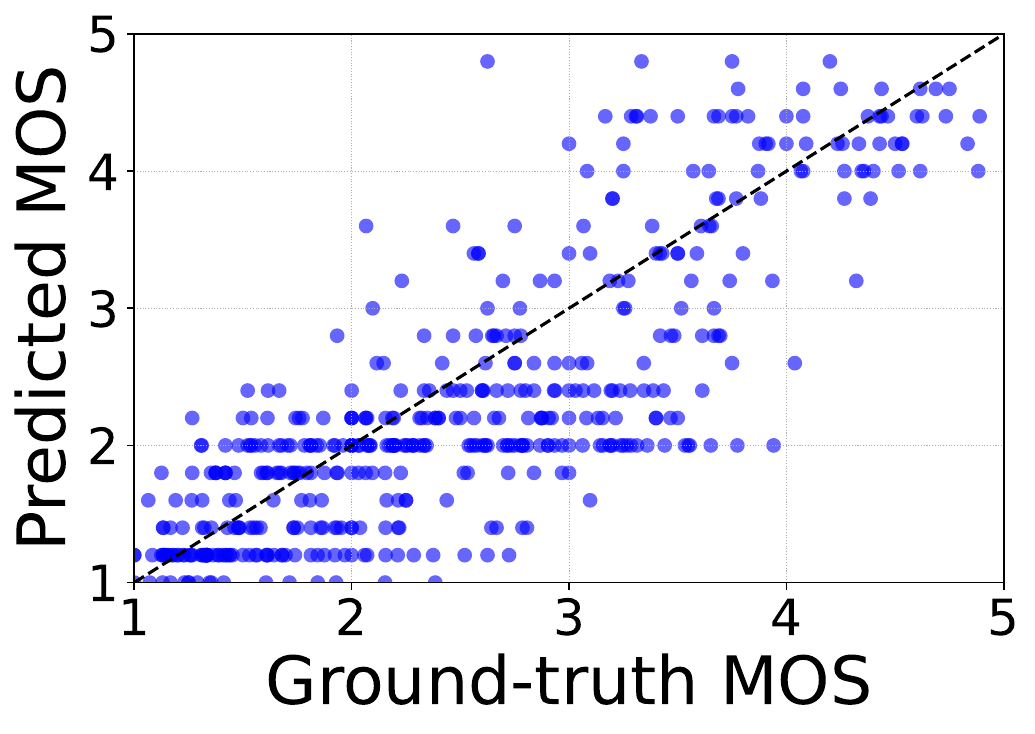}
    \caption{Full-ref AST (frozen)}
    \label{fig:mos-scatter-fullref-ast-fr}
  \end{subfigure}
  % Third image
  \begin{subfigure}[b]{0.19\textwidth}
    \centering
    \includegraphics[width=\linewidth]{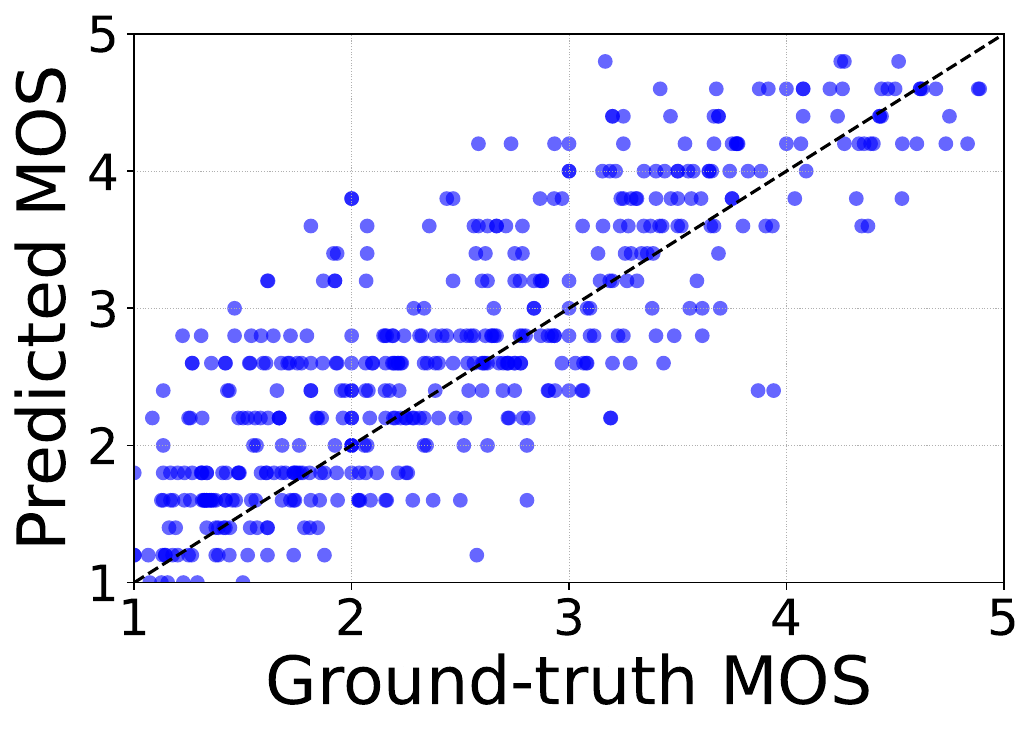}
    \caption{Full-ref Whisper}
    \label{fig:mos-scatter-fullref-whisper}
  \end{subfigure}
  % Fourth image
  \begin{subfigure}[b]{0.19\textwidth}
    \centering
    \includegraphics[width=\linewidth]{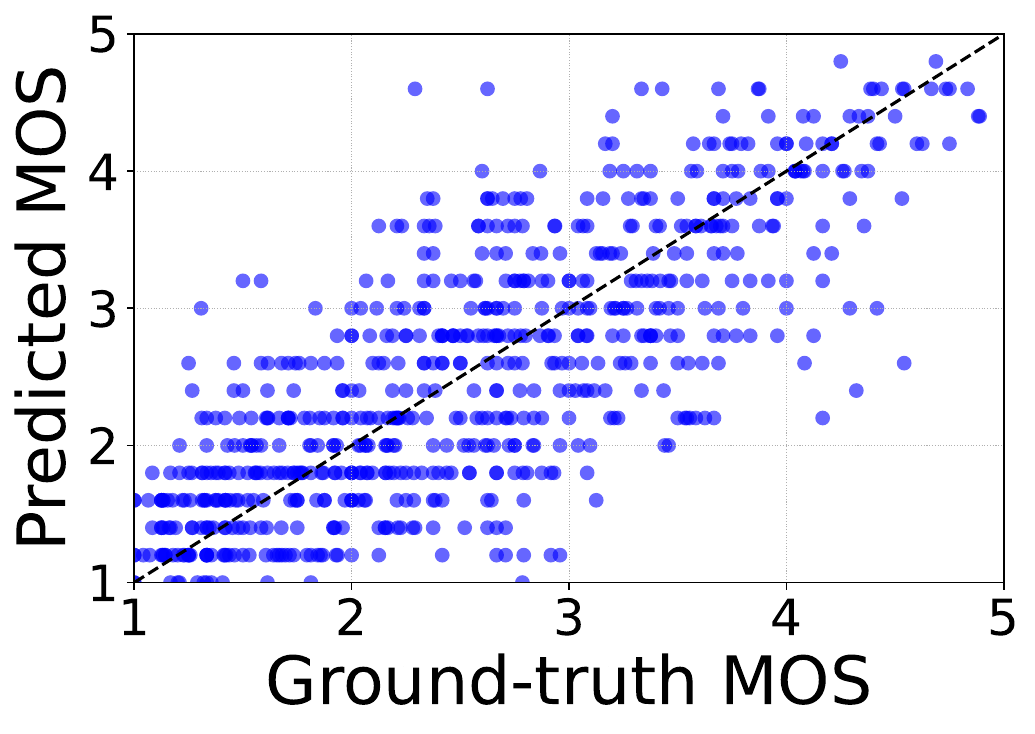}
    \caption{No-ref AST (finetuned)}
    \label{fig:mos-scatter-noref-ast-ft}
  \end{subfigure}
  % Fifth image
  \begin{subfigure}[b]{0.19\textwidth}
    \centering
    \includegraphics[width=\linewidth]{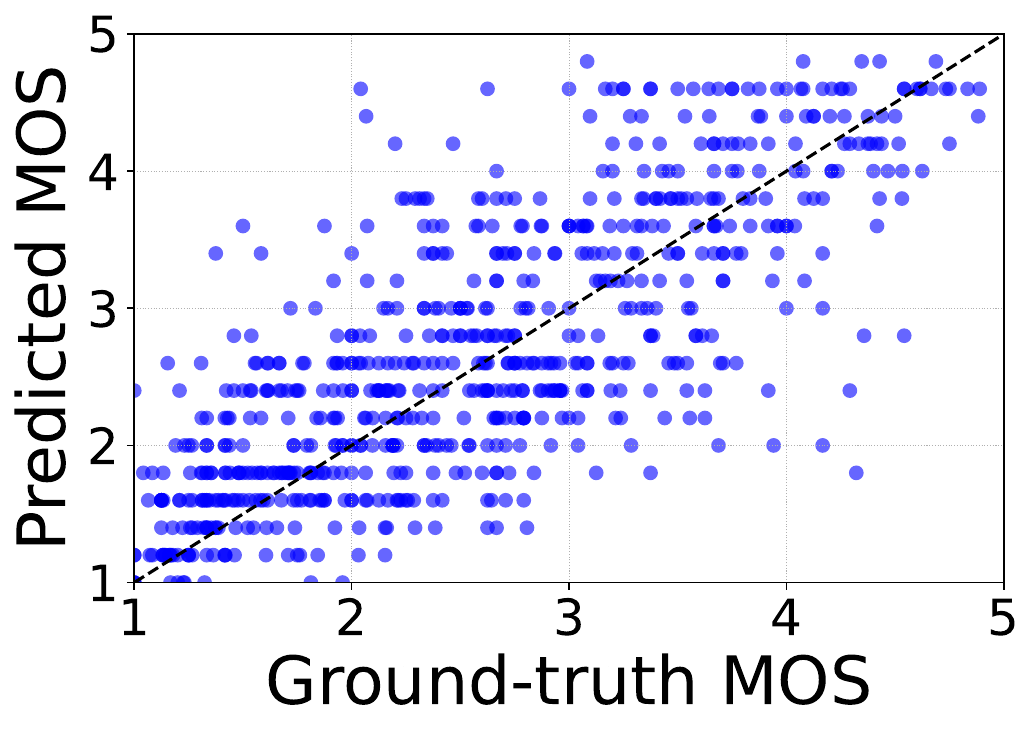}
    \caption{No-ref AST (frozen)}
    \label{fig:mos-scatter-noref-ast-fr}
  \end{subfigure}

  \caption{Scatter plots of predicted versus ground-truth MOS of the five configurations for the \textit{MOS-numeric} task.}
  \label{fig:mos-scatter-all}
\end{figure*}

\begin{figure*}[t]
  \centering
  % First image
  \begin{subfigure}[b]{0.19\textwidth}
    \centering
    \includegraphics[width=\linewidth]{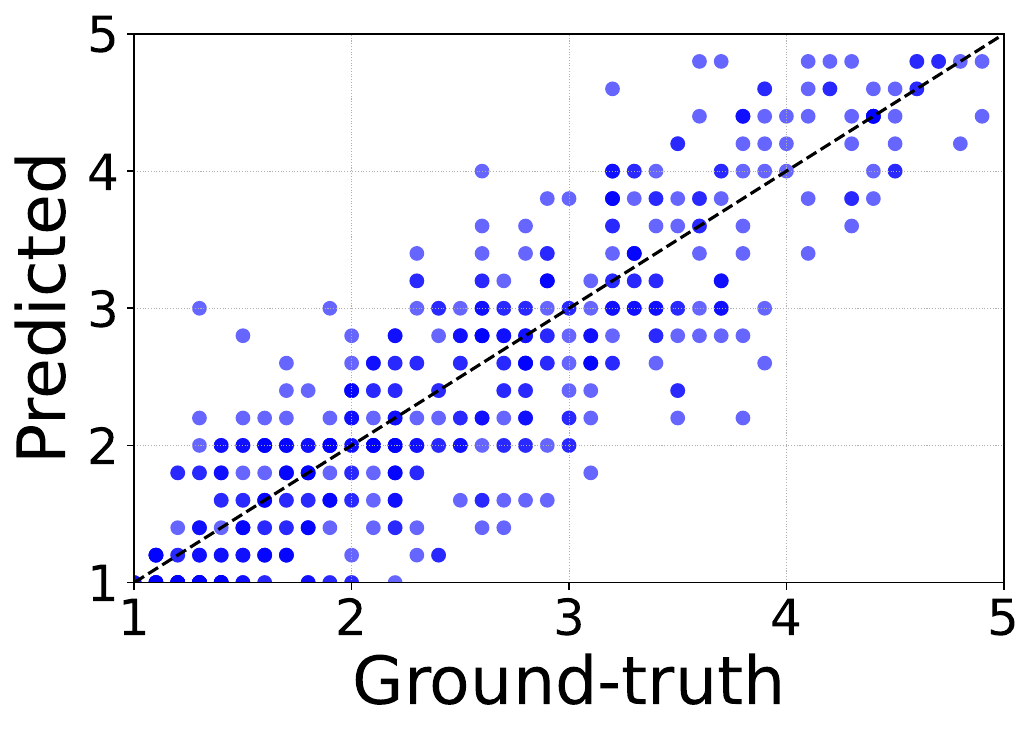}
    \caption{MOS}
    \label{fig:mos-multidim}
  \end{subfigure}
  % Second image
  \begin{subfigure}[b]{0.19\textwidth}
    \centering
    \includegraphics[width=\linewidth]{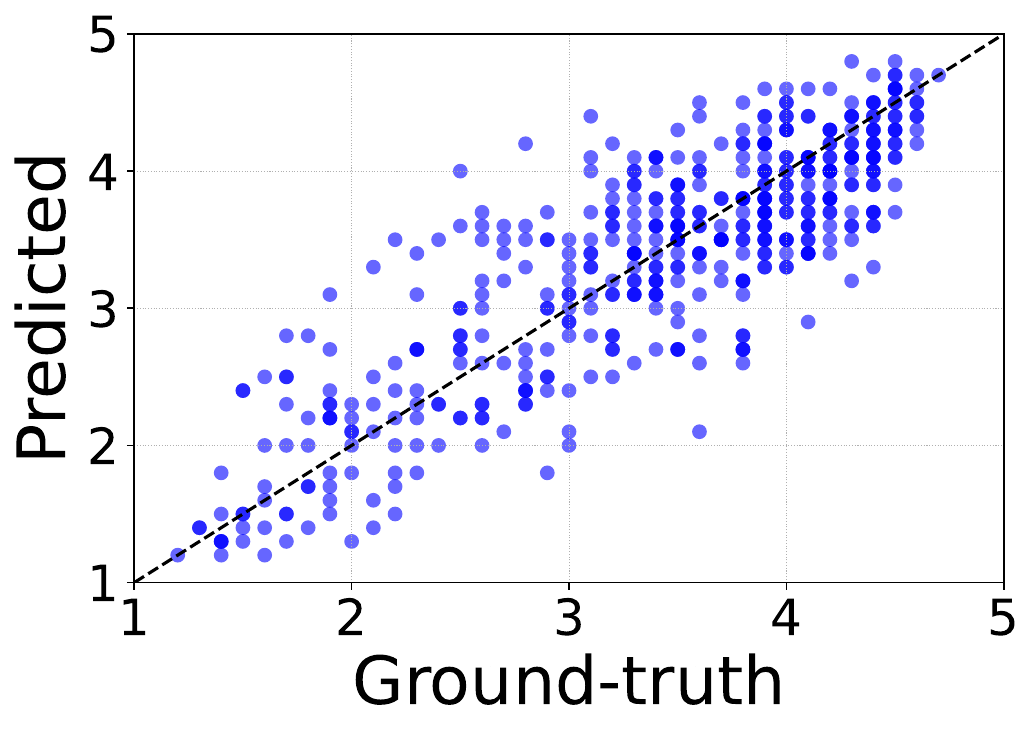}
    \caption{Noisiness}
    \label{fig:nois-multidim}
  \end{subfigure}
  % Third image
  \begin{subfigure}[b]{0.19\textwidth}
    \centering
    \includegraphics[width=\linewidth]{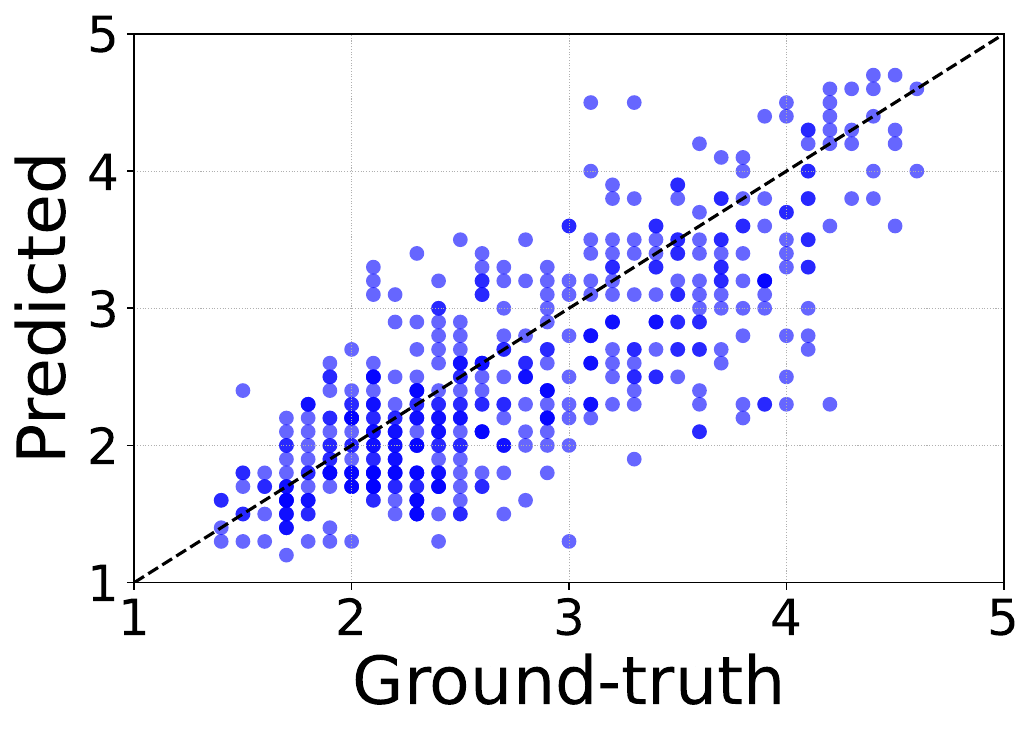}
    \caption{Coloration}
    \label{fig:col-multidim}
  \end{subfigure}
  % Fourth image
  \begin{subfigure}[b]{0.19\textwidth}
    \centering
    \includegraphics[width=\linewidth]{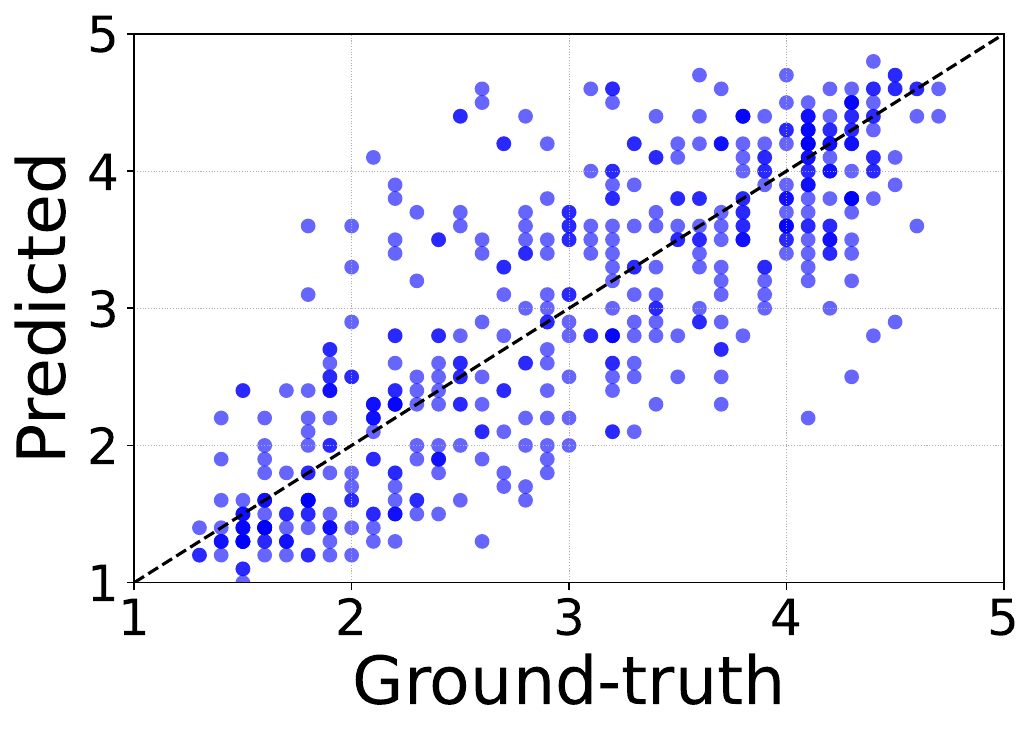}
    \caption{Discontinuity}
    \label{fig:dis-multidim}
  \end{subfigure}
  % Fifth image
  \begin{subfigure}[b]{0.19\textwidth}
    \centering
    \includegraphics[width=\linewidth]{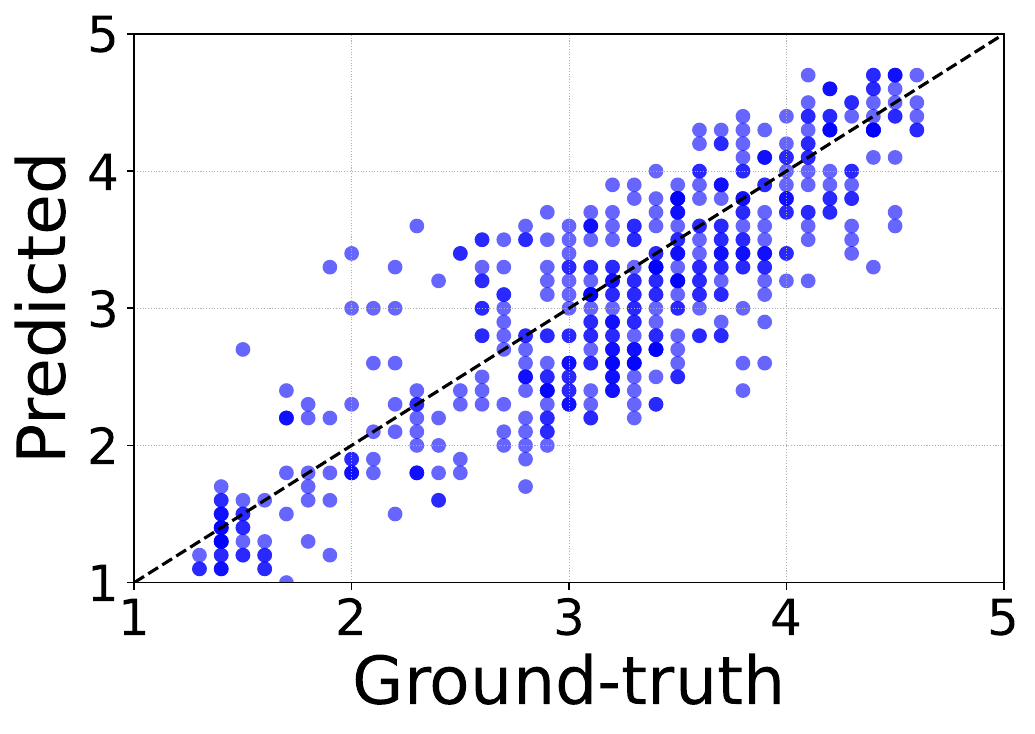}
    \caption{Loudness}
    \label{fig:loud-multidim}
  \end{subfigure}

  \caption{Scatter plots of predicted versus ground-truth MOS and the four dimensions for the \textit{Multi-dim} task using Full-ref AST~(finetuned).}
  \label{fig:multidim}
\end{figure*}

\begin{figure*}[t]
  \centering
  % First image
  \begin{subfigure}[b]{0.19\textwidth}
    \centering
    \includegraphics[width=\linewidth]{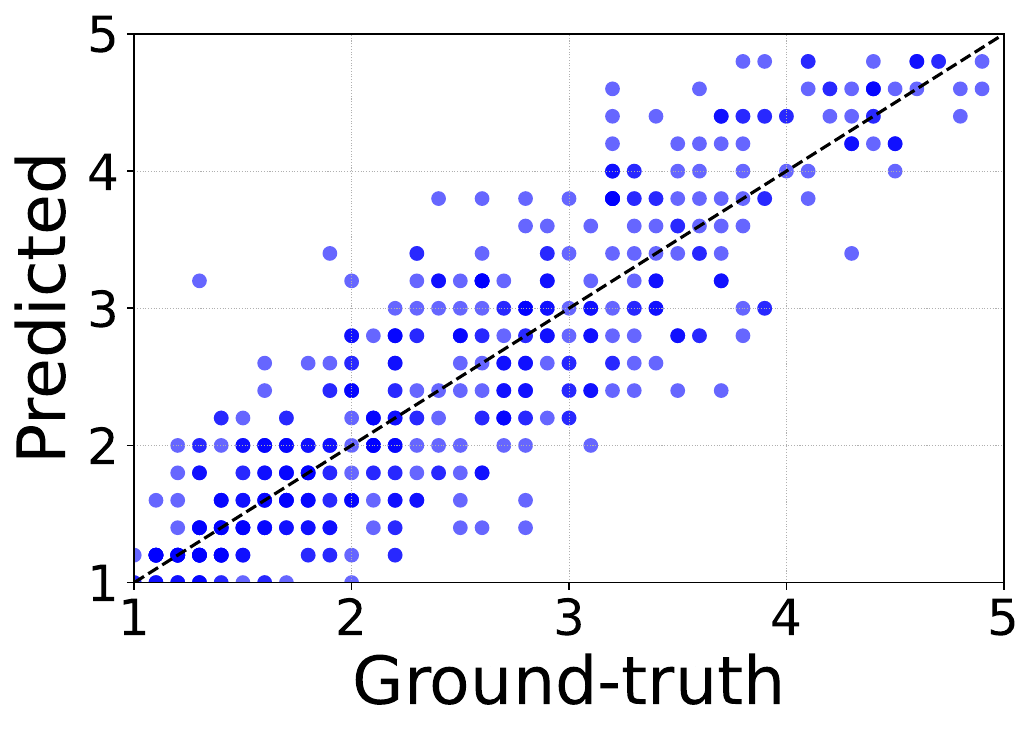}
    \caption{MOS}
    \label{fig:mos-exp}
  \end{subfigure}
  % Second image
  \begin{subfigure}[b]{0.19\textwidth}
    \centering
    \includegraphics[width=\linewidth]{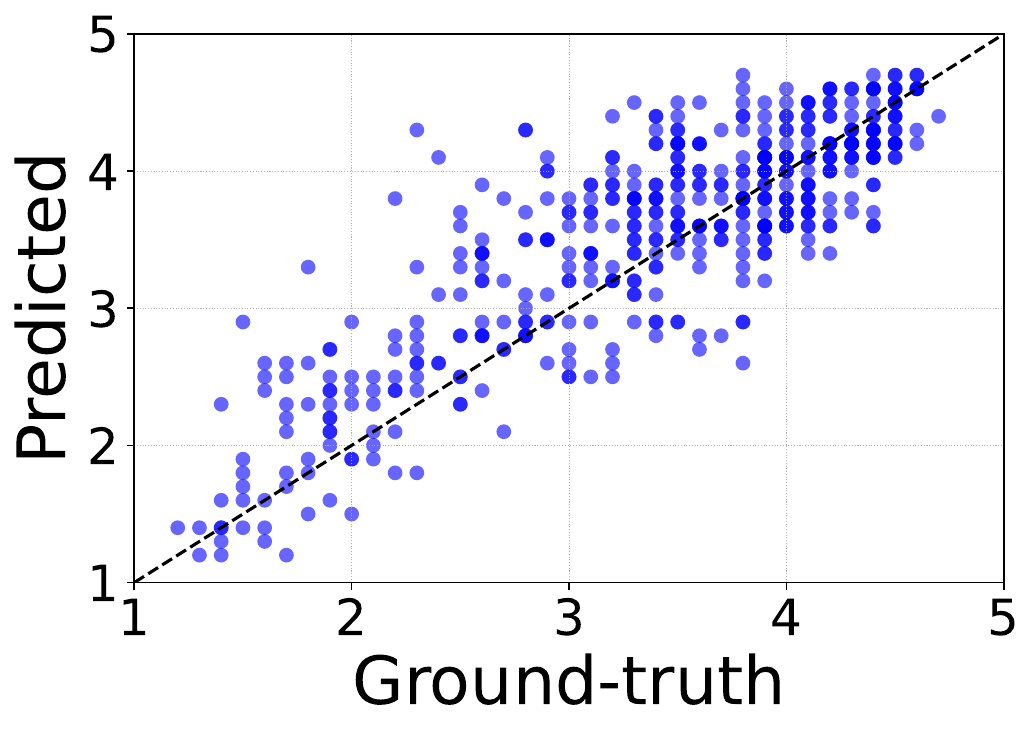}
    \caption{Noisiness}
    \label{fig:nois-exp}
  \end{subfigure}
  % Third image
  \begin{subfigure}[b]{0.19\textwidth}
    \centering
    \includegraphics[width=\linewidth]{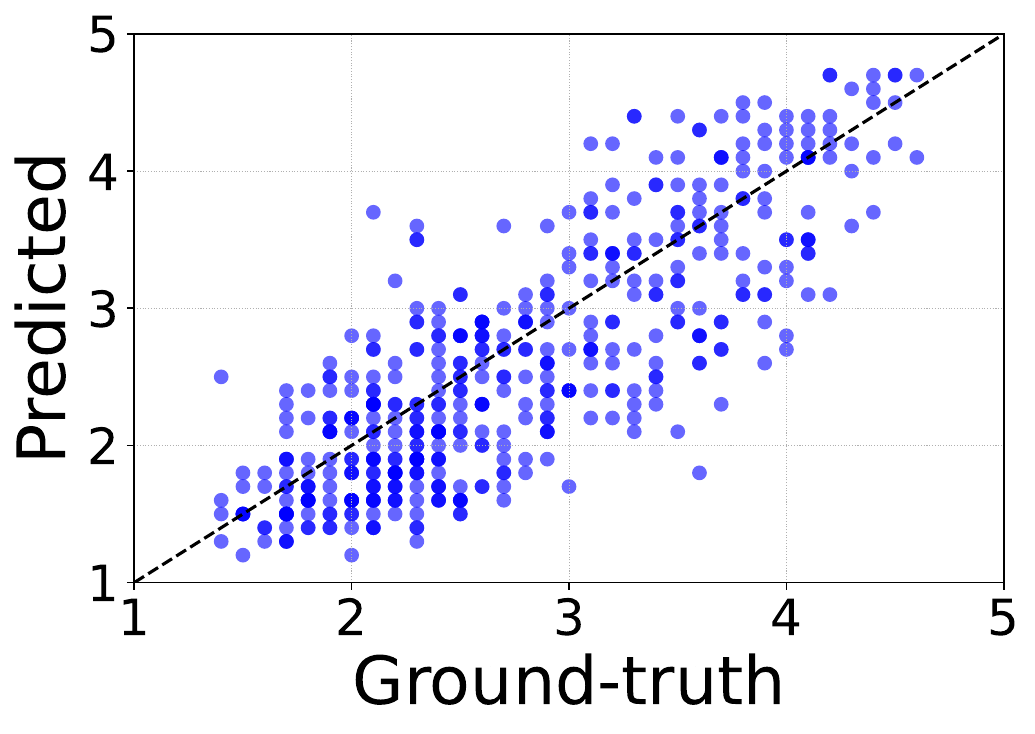}
    \caption{Coloration}
    \label{fig:col-exp}
  \end{subfigure}
  % Fourth image
  \begin{subfigure}[b]{0.19\textwidth}
    \centering
    \includegraphics[width=\linewidth]{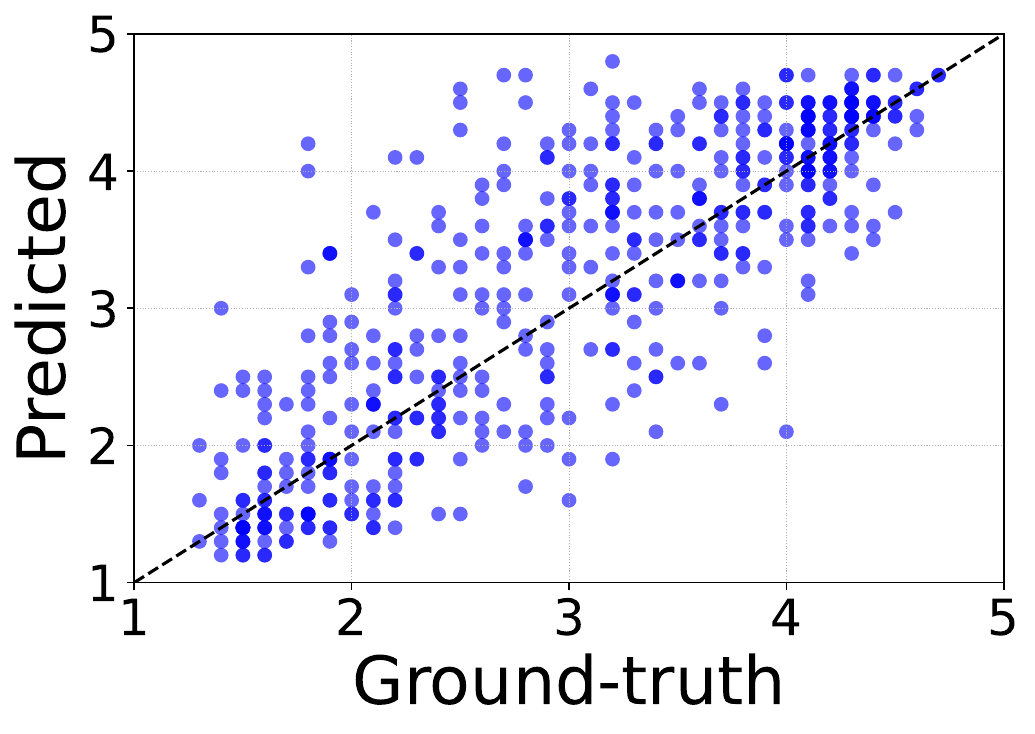}
    \caption{Discontinuity}
    \label{fig:dis-exp}
  \end{subfigure}
  % Fifth image
  \begin{subfigure}[b]{0.19\textwidth}
    \centering
    \includegraphics[width=\linewidth]{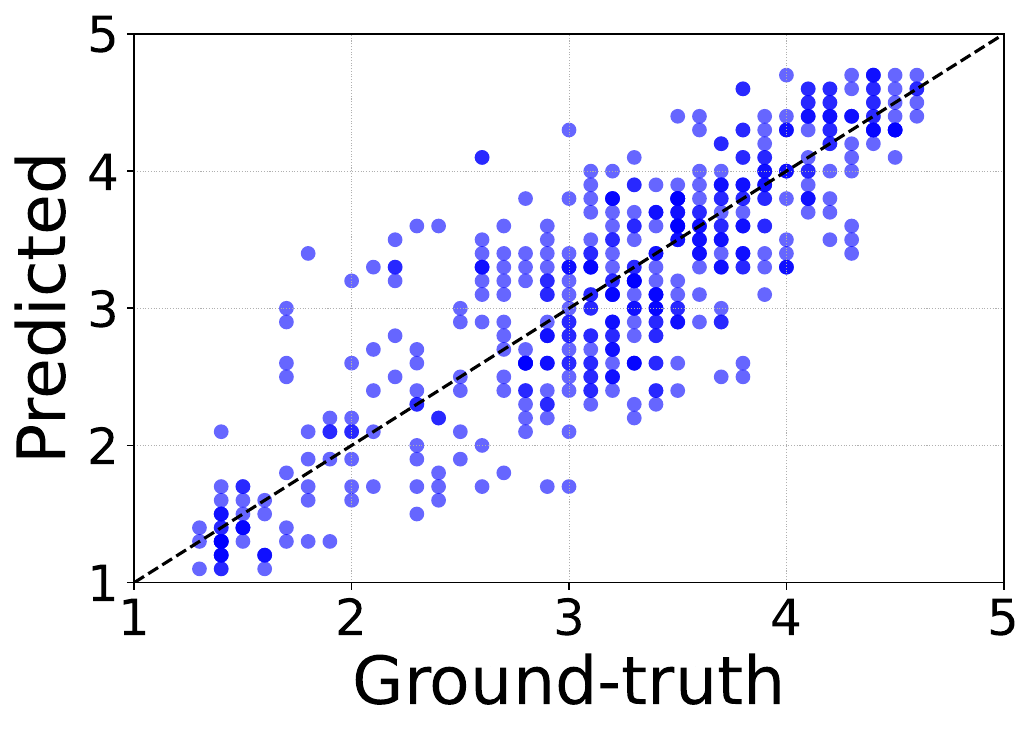}
    \caption{Loudness}
    \label{fig:loud-exp}
  \end{subfigure}

  \caption{Scatter plots of predicted versus ground-truth MOS and the four dimensions for the \textit{Explanatory} task using Full-reference AST~(finetuned).}
  \label{fig:exp}
\end{figure*}

\section{Evaluation}

\subsection{Experimental Setup}

We evaluate \sys on the official NISQA test partitions in five configurations that mirror the training setups: (i) \textbf{Full-reference AST (frozen)}, which uses an AST audio encoder with all backbone weights frozen; (ii) \textbf{Full-reference AST (finetuned)}, where we additionally unfreeze the AST attention query projection parameters; (iii) \textbf{Full-reference Whisper (frozen)}, which replaces AST with a Whisper encoder; (iv) \textbf{No-reference AST (frozen)}, a single-ended setting without a clean reference; and (v) \textbf{No-reference AST (finetuned)}, the corresponding single-ended finetuned variant.

For each test clip, we instantiate the five QA task families described in Table~\ref{tab:qa-tasks}: (1) \emph{MOS-numeric}, which elicits an overall MOS on a 1--5 scale; (2) \emph{Dim-numeric}, which targets 1--5 ratings for noisiness, coloration, discontinuity, and loudness; (3) \emph{Dim-categorical}, which asks for verbal quality categories that can be mapped back to numeric scores; (4) \emph{Multi-dim}, which requests a single answer summarizing the overall MOS and all four dimensions; and (5) \emph{Explanatory}, which asks for a short natural-language justification that includes an overall MOS. For each task and model configuration, we generate one answer per test clip using the model’s standard text generation interface, starting from the same system prompt and QA template bank as in training. The resulting free-form outputs are then parsed with a deterministic set of regular-expression and keyword rules to recover numeric scores; clips for which no numeric value can be extracted are omitted from quantitative score-based metrics but are retained for qualitative analysis of the model’s reasoning behavior.

\subsection{Evaluation Metrics}

We evaluate predictive performance using standard regression and ranking metrics between the model outputs and the ground-truth scores: mean absolute error (MAE), root mean squared error (RMSE), Pearson correlation coefficient ($r$), and Spearman rank correlation coefficient ($\rho$). These metrics are computed separately for each task family and for each perceptual dimension, allowing us to assess both overall MOS prediction and dimension-wise quality estimation.

\subsection{Overall MOS Prediction}

Table~\ref{tab:mos-numeric} summarizes performance on the \emph{MOS-numeric} task, in which the model is asked to predict an overall MOS on the 1--5 scale. The full-reference AST finetuned configuration achieves the best performance, with an MAE of $0.41$ and a Pearson correlation of $r = 0.86$ on the test clips, clearly outperforming its frozen counterpart and all no-reference variants. The full-reference Whisper model ranks second, with a slightly higher MAE but comparable correlation. In the single-ended setting, finetuning the AST encoder yields a modest but consistent improvement over the frozen baseline, reducing MAE from $0.51$ to $0.49$ and increasing the correlation from $r = 0.77$ to $r = 0.78$. We show the scatter plots of MOS scores for all five \sys settings in Figure~\ref{fig:mos-scatter-all}.

\begin{table}
  \centering
  \small
  \caption{Overall MOS prediction on NISQA test clips for the MOS-numeric task.}
  \label{tab:mos-numeric}
  \begin{tabular}{lcccc}
    \toprule
    Setting & MAE & RMSE & $r$ & $\rho$\\
    \midrule
    Full-ref, AST (finetuned) & \textbf{0.41} & \textbf{0.52} & \textbf{0.86} & \textbf{0.84} \\
    Full-ref, Whisper (frozen)& 0.47 & 0.61 & 0.83 & 0.81 \\
    Full-ref, AST (frozen)    & 0.48 & 0.62 & 0.82 & 0.8 \\
    No-ref, AST (finetuned)   & 0.49 & 0.63 & 0.78 & 0.77 \\
    No-ref, AST (frozen)      & 0.51 & 0.67 & 0.77 & 0.77 \\
    \bottomrule
  \end{tabular}
\end{table}

The \emph{Explanatory} task, in which the model must produce a short textual rationale that includes an MOS, exhibits very similar quantitative behavior: the full-reference AST finetuned model again attains an MAE of $0.4$ and a Pearson correlation of $r = 0.88$ for MOS, indicating that generating free-form explanations does not noticeably degrade score prediction. Single-ended models achieve MAEs in the range $0.47$--$0.53$ with correlations between $r = 0.77$ and $r = 0.80$, which is competitive with state-of-the-art DNN-based objective metrics while additionally providing rich natural-language justifications.

\subsection{Dimension-wise Quality Prediction}

We next analyze the \emph{Dim-numeric} task, where each question targets a single perceptual dimension (noisiness, coloration, discontinuity, loudness) and the model must return a 1--5 score. Table~\ref{tab:dim-numeric-mae} reports mean absolute error (MAE) for all five model configurations, and Table~\ref{tab:dim-numeric-r} reports the corresponding Pearson correlations. Across all dimensions, the full-reference AST finetuned model attains the strongest or near-strongest correlations, achieving $r=0.8$ for noisiness and $r=0.82$ for discontinuity with MAEs around $0.41$–$0.56$ on a 1–5 scale. Relative to the frozen AST baseline, finetuning substantially improves discontinuity ($r$ from $0.43$ to $0.82$, MAE from $0.78$ to $0.47$) and yields consistent gains for noisiness and loudness as well. The full-reference Whisper model performs worse than finetuned AST on all four dimensions, particularly for coloration and loudness, suggesting that AST’s time–frequency representation is better aligned with the perceptual attributes of audio.

In the single-ended setting, both no-reference AST models achieve strong dimension-wise correlations despite only seeing the degraded signal. The finetuned variant reaches $r=0.69$–$0.76$ across all four dimensions with MAEs in the $0.44$–$0.59$ range, closely matching the full-reference AST finetuned model for noisiness, coloration, and loudness. The frozen no-reference AST model is consistently weaker, especially for discontinuity (MAE $0.70$, $r=0.64$), but remains competitive overall. These results indicate that, while access to a clean reference yields the best performance, a suitably tuned single-ended model can still deliver reliable dimension-wise predictions.

\begin{table}[t]
  \centering
  \small
  \caption{MAE for numeric dimension prediction (Dim-numeric task).}
  \label{tab:dim-numeric-mae}
  \begin{tabular}{lcccc}
    \toprule
    Setting & Noise & Color & Disc. & Loud \\
    \midrule
    Full-ref, AST (finetuned) & \textbf{0.43} & 0.56 & \textbf{0.47} & \textbf{0.41} \\
    Full-ref, AST (frozen)    & 0.51 & 0.66 & 0.78 & 0.78 \\
    Full-ref, Whisper (frozen)& 0.69 & 0.69 & 0.78 & 0.77 \\
    No-ref, AST (finetuned)   & 0.58 & \textbf{0.48} & 0.59 & 0.44 \\
    No-ref, AST (frozen)      & 0.53 & 0.55 & 0.70 & 0.60 \\
    \bottomrule
  \end{tabular}
\end{table}

\begin{table}[t]
  \centering
  \small
  \caption{Pearson correlation ($r$) for numeric dimension prediction (Dim-numeric task).}
  \label{tab:dim-numeric-r}
  \begin{tabular}{lcccc}
    \toprule
    Setting & Noise & Color & Disc. & Loud \\
    \midrule
    Full-ref, AST (finetuned) & \textbf{0.8} & 0.63 & \textbf{0.82} & \textbf{0.79} \\
    Full-ref, AST (frozen)    & 0.73 & 0.54 & 0.43 & 0.53 \\
    Full-ref, Whisper (frozen)& 0.60 & 0.44 & 0.48 & 0.43 \\
    No-ref, AST (finetuned)   & 0.71 & \textbf{0.69} & 0.75 & 0.76 \\
    No-ref, AST (frozen)      & 0.70 & 0.64 & 0.64 & 0.56 \\
    \bottomrule
  \end{tabular}
\end{table}

We also evaluate a \emph{Dim-categorical} task, where the model outputs coarse verbal categories (e.g., very bad, bad, fair, good, excellent) that are mapped back to discrete scores. Table~\ref{tab:dim-categ-r} reports Pearson correlations for all configurations. The finetuned AST model yields near-complete coverage for all dimensions and achieves reasonable correlations (e.g., $r=0.82$ for noisiness and $r=0.8$ for coloration). However, other configurations struggle in many dimensions, mainly failing to produce explanation in some cases, resulting in non-parsable format for metric comparison. The Whisper model attains its strongest categorical performance on coloration ($r=0.75$) but struggles on loudness. The no-reference AST models show mixed behavior: the finetuned variant attains $r=0.70$ for discontinuity but lower correlations for noisiness and loudness, whereas the frozen variant offers more balanced but generally weaker performance. Overall, the categorical task confirms that \sys can produce qualitatively sensible verbal ratings, but also highlights the sensitivity of discrete-label parsing to template adherence.

\begin{table}[t]
  \centering
  \small
  \caption{Pearson correlation ($r$) for categorical dimension prediction (Dim-categorical task).}
  \label{tab:dim-categ-r}
  \begin{tabular}{lcccc}
    \toprule
    Setting & Noise & Color & Disc. & Loud \\
    \midrule
    Full-ref, AST (finetuned) & \textbf{0.82} & \textbf{0.8} & \textbf{0.77} & \textbf{0.71} \\
    Full-ref, AST (frozen)    & 0.73 & 0.62 & 0.46 & 0.41 \\
    Full-ref, Whisper (frozen)& 0.61 & 0.75 & 0.46 & 0.32 \\
    No-ref, AST (finetuned)   & 0.47 & 0.62 & 0.70 & 0.41 \\
    No-ref, AST (frozen)      & 0.63 & 0.44 & 0.26 & 0.45 \\
    \bottomrule
  \end{tabular}
\end{table}

\begin{table}[t]
  \centering
  \small
  \setlength{\tabcolsep}{4pt}  % shrink horizontal padding a bit
  \caption{Pearson correlation ($r$) for MOS and dimensions in the Multi-dim task.}
  \label{tab:multi-dim-r}
  \begin{tabular}{p{2.7cm}ccccc}
    \toprule
    Setting & MOS & Noise & Color & Disc. & Loud \\
    \midrule
    Full-ref, AST \\(finetuned)  & \textbf{0.88} & \textbf{0.85} & \textbf{0.79} & \textbf{0.82} & \textbf{0.86} \\
    \hline
    Full-ref, AST \\(frozen)     & 0.84 & 0.86 & 0.74 & 0.46 & 0.60 \\
    \hline
    Full-ref, Whisper \\(frozen) & 0.82 & 0.75 & 0.38 & 0.44 & 0.69 \\
    \hline
    No-ref, AST \\(finetuned)    & 0.76 & 0.74 & 0.73 & 0.72 & 0.77 \\
    \hline
    No-ref, AST \\(frozen)       & 0.79 & 0.78 & 0.72 & 0.63 & 0.66 \\
    \hline
    \bottomrule
  \end{tabular}
\end{table}

\begin{table*}[t]
  \centering
  \small
  \caption{Example question--answer pairs for the five task families in our evaluation. For each task, we show real test clips with the corresponding question, ground-truth answer derived from NISQA labels, and the model-generated answer.}
  \label{tab:qa-examples}
  \begin{tabular}{|p{1.5cm}|p{5.5cm}|p{4.5cm}|p{4.5cm}|}
    \toprule
    \textbf{Task} & \textbf{Question} & \textbf{Ground-truth answer} & \textbf{Generated answer} \\
    \midrule

    MOS-numeric &
    On a scale from 1 (very bad) to 5 (excellent), what is the overall listening quality of the degraded audio? &
    4.3 &
    4.0 \\
    \hline 

    MOS-numeric &
    Rate the overall mean opinion score (MOS) of the degraded speech using a 1--5 scale. &
    I would rate the overall MOS as 1.2 out of 5. &
    1.4 \\
    \hline

    Dim-numeric (Noise) &
    On a 1–5 scale, how would you score the noisiness of the degraded signal relative to the clean reference? &
    4.3 on the 1--5 scale. &
    4.2 on the 1--5 scale. \\
    \hline

    Dim-numeric (Loudness) &
    Rate the loudness quality of the degraded speech compared to the reference on a 1--5 scale (higher means more natural loudness). &
    4.5 on the 1--5 scale. &
    4.6 on the 1--5 scale. \\
    \hline

    Dim-categorical (Noise) &
    How is the degraded audio in terms of background noisiness? Use categories: very bad, poor, fair, good, or excellent. &
    The noisiness quality is good, about 4.3 out of 5. &
    I would say the noisiness is good ($\approx$4.1/5). \\
    \hline

    Dim-categorical (Discontinuity) &
    How would you describe temporal discontinuities in the degraded audio (clicks, dropouts, glitches)? Use very bad, poor, fair, good, or excellent. &
    The discontinuity quality is good, about 4.1/5. &
    The discontinuity quality is good, about 3.9/5. \\
    \hline

    Multi-dim &
    Provide a quality assessment of the degraded speech compared to the reference, including overall MOS and the four dimensions: noisiness, coloration, discontinuity, and loudness. Give scores between 1 and 5. &
    Overall MOS: 4.3. Noisiness: 4.3. Coloration: 4.2. Discontinuity: 4.4. Loudness quality: 4.4. &
    I would assign the following scores (1--5): overall MOS = 4.2, noisiness = 4.1, coloration = 3.6, discontinuity = 4.1, loudness = 4.3. \\
    \hline

    Explanatory &
    Explain the main causes of quality degradation in the degraded speech compared to the reference, then provide an overall MOS score between 1 and 5. &
    The degraded audio suffers from Anchor: Fullband clean. Considering the combined effects on noisiness ($\approx$4.3), coloration ($\approx$4.2), discontinuity ($\approx$4.4), and loudness quality ($\approx$4.4), I would give an overall MOS of 4.3. &
    The degraded audio suffers from simulated. Considering the combined effects on noisiness ($\approx$4.2), coloration ($\approx$4.2), discontinuity ($\approx$4.5), and loudness quality ($\approx$4.3), I would give an overall MOS of 4.2. \\
    \hline

    \bottomrule
  \end{tabular}
\end{table*}

Finally, we consider the \emph{Multi-dim} task, in which a single answer must summarize the overall MOS and all four dimensions. Table~\ref{tab:multi-dim-r} reports Pearson correlations for each dimension under this setting and Figure~\ref{fig:multidim} shows the scatter plots of MOS and the four perceptual dimensions. The finetuned full-reference AST model achieves the best multi-dimensional performance for most aspects, with $r=0.88$ for MOS, $0.85$ for noisiness, and $0.79$ for coloration. The frozen full-reference AST model, while still strong, shows lower discontinuity ($r=0.46$ for disc) and loudness ($r=0.6$). The Whisper variant again lags behind AST, particularly for coloration and discontinuity.

In the no-reference regime, the finetuned AST model attains MOS correlation $r=0.76$ with MAE $0.50$, and dimension-wise correlations between $0.72$ and $0.77$, demonstrating that the model can jointly reason about multiple quality dimensions from degraded audio alone. The frozen no-reference AST baseline slightly trails in MOS ($r=0.79$, MAE $0.53$) but surprisingly matches or exceeds finetuned performance on some dimensions, likely due to more consistent numeric formatting in its summaries. Together, these results suggest a trade-off between highly structured multi-dim answers (which favor frozen encoders) and more flexible, free-form responses (which benefit from encoder finetuning but are harder to parse exhaustively).

Overall, the expanded dimension-wise analysis shows that \sys can reliably predict both numeric and categorical ratings across all four perceptual dimensions, with full-reference AST finetuning providing the strongest performance and single-ended AST models remaining competitive. The Multi-dim task further demonstrates that the model can produce coherent joint assessments of MOS and dimensions, albeit with an accuracy–formatting trade-off between finetuned and frozen encoders.

\subsection{Effect of Reference Signal and Encoder Choice}

Comparing full-reference and no-reference settings highlights the value of clean references. For overall MOS, adding a reference consistently reduces MAE by about $0.08$--$0.10$ and boosts Pearson correlation by $0.05$--$0.08$, both for frozen and finetuned encoders. The gap is especially pronounced for clips with severe distortions, where the single-ended model must disentangle channel effects from degradations purely from the degraded waveform. However, the strong performance of the no-reference finetuned AST variant (MAE $\approx 0.49$ and $r\approx 0.78$) suggests that \sys remains competitive in deployment scenarios where reference signals are unavailable.

Encoder choice also plays a significant role in overall performance. Among the full-reference configurations, the finetuned AST model provides the strongest predictions for both MOS and the perceptual dimensions, with the frozen AST variant trailing only slightly behind. Whisper, although a highly capable speech encoder, consistently lags behind AST on several dimension-wise metrics. A plausible explanation is that Whisper is explicitly trained to be robust to background noise and channel artifacts in order to preserve linguistic content, and thus learns to suppress precisely those degradations that are most salient for perceived quality. By contrast, AST is pretrained on AudioSet, which spans a broad mixture of speech, noise, and environmental sounds, encouraging the encoder to retain relative energy levels and fine-grained spectral cues that are more diagnostic of noisiness, coloration, discontinuity, and loudness. Comparing frozen and finetuned variants further indicates that modest encoder finetuning yields consistent gains across all tasks, effectively specializing the audio backbone for speech quality assessment rather than generic recognition.

\subsection{Textual Outputs and Interpretability}

Beyond scalar scores, \sys also produces natural-language answers for every question, which in principle can reveal the perceived causes of degradation. In the \emph{Explanatory} setting, the model is prompted to output an overall MOS together with a short textual justification. The corresponding scatter plots for this setting are shown in Figure~\ref{fig:exp}, and examples of our template-based question–answer pairs, along with model-generated responses, are provided in Table~\ref{tab:qa-examples}. While the predicted MOS values and their explanations are generally consistent with the ground truth (as reflected by the quantitative metrics in the scatter plots), we observe that the model often resorts to highly similar phrasing when describing the audio condition (e.g., repeatedly commenting on SNR or background noise in a stereotyped way). We hypothesize that this behavior is partly induced by the specificity of our templated conditions, and that relaxing the templates to encourage more free-form justifications could yield richer, more diverse explanations.

\begin{table}[t]
  \centering
  \small
  \setlength{\tabcolsep}{4pt}  % reduce horizontal padding
  \caption{Text-based similarity metrics between ground-truth and model-generated answers on the Dim-categorical task.}
  \label{tab:text-metrics}
  \begin{tabular}{p{2.2cm}ccc}
    \toprule
    Setting & sacreBLEU~($\uparrow$) & BLEU~($\uparrow$) & ROUGE-L~($\uparrow$) \\
    \midrule
    Full-ref, AST \\(finetuned)  & \textbf{54.78} & \textbf{0.55} & \textbf{0.78} \\
    \hline
    Full-ref, AST \\(frozen)     &  5.74          & 0.11          & 0.29          \\
    \hline
    Full-ref, Whisper \\(frozen) &  2.72          & 0.01          & 0.21          \\
    \hline
    No-ref, AST \\(finetuned)    & 12.64          & 0.04          & 0.34          \\
    \hline
    No-ref, AST \\(frozen)       & 12.10          & 0.08          & 0.27          \\
    \bottomrule
  \end{tabular}
\end{table}

On the Dim-categorical task, text-based metrics clearly separate the full-reference finetuned AST model from all other configurations. As shown in Table~\ref{tab:text-metrics}, the finetuned full-reference AST system attains a corpus-level sacreBLEU of $54.78$, a mean sentence-level BLEU of $0.55$, and a mean ROUGE-L F1 of $0.78$, indicating that its rationales closely track the ground-truth templates in both wording and structure. In contrast, the frozen full-reference AST baseline drops to sacreBLEU $5.74$, BLEU mean $0.11$, and ROUGE-L mean $0.29$, while the frozen Whisper encoder performs even worse (sacreBLEU $2.72$, BLEU mean $0.01$, ROUGE-L mean $0.21$), reflecting frequent deviations from the expected explanatory patterns. The single-ended AST models lie in between: despite having access only to the degraded signal, they achieve moderate ROUGE-L overlap (mean $0.34$ for the finetuned variant and $0.27$ for the frozen one) but relatively low BLEU, suggesting that they often convey qualitatively similar content using more variable phrasing. Overall, these results show that finetuning the AST-based full-reference system not only improves numeric quality prediction but also yields explanations that are linguistically much closer to the intended reference rationales.

\section{Conclusion}

In this work, we introduced \sys, a multimodal speech quality QA system that casts objective assessment as a natural-language interaction task and uses a large language model as an “expert listener.” Developing and evaluating on NISQA, we constructed on-the-fly question–answer pairs that cover overall MOS and four perceptual dimensions, and coupled them with either AST- or Whisper-based encoders in both single-ended and double-ended configurations. Across five model variants, \sys achieves competitive MOS prediction compared to strong DNN baselines, with the full-reference AST model reaching low MAE and high correlation while simultaneously supporting text-based queries and explanations; even the no-reference models attain robust dimension-wise performance using only degraded audio. Looking forward, we plan to enrich the explanatory supervision so that the model produces more detailed, artifact-specific rationales; to systematically study profile-conditioned MOS distributions that emulate diverse user populations; to extend training and evaluation to broader corpora, languages, and real-time streaming conditions; and to integrate human-in-the-loop feedback, including targeted listening tests, to further align \sys with subjective perception in real-world scenarios. In parallel, we aim to explore tighter joint optimization of the audio encoder and language model, including multi-task objectives that explicitly balance numeric accuracy and explanation quality. Finally, by releasing code, prompts, and evaluation protocols, we hope \sys can serve as a reproducible benchmark for future work at the intersection of speech quality modeling and language-based interaction.

{
    %\balance
    \small
    \bibliographystyle{ieeenat_fullname}
    \bibliography{references}
}

\end{document}